\documentclass[aps,prl,groupedaddress,nofootinbib]{revtex4-2}  
\usepackage{graphicx}  
\usepackage{dcolumn}   
\usepackage{bm}        
\usepackage{amssymb}   
\usepackage{slashed}
\usepackage{color}
\usepackage{lipsum}
\usepackage{textcomp}
\usepackage{appendix}
\begin{document}	
 \title{  Inclusive beauty-charmed baryons decay $ \Xi_{bcq} \to  \Xi_{ccq} +X$ }	
\author{Guo-He Yang$^{1}$\footnote{ygh@hnit.edu.cn} }

\address{
$^1$College of science, Hunan Institute of  Technology, Hunan 421002, China }
\date{\today}
\providecommand{\keywords}[1]{\textbf{\textit{keywords---}} #1}	

\begin{abstract} 
We investigate the inclusive weak decay $\Xi_{bcq} \to \Xi_{ccq} + X$ ($q=u,d,s$) within the framework of the non-relativistic potential quark model. Treating the heavy diquarks $\Phi_{bc}$ and $\Phi_{cc}$ as compact color anti-triplet bound states, we calculate the transition form factors via the overlap integral of Schrödinger wave functions derived from the Cornell potential. The decay rates for four dominant channels—$\Phi_{bc} \to \Phi_{cc} + \bar{c}s$, $\Phi_{bc} \to \Phi_{cc} + \bar{u}s$, $\Phi_{bc} \to \Phi_{cc} + e/\mu + \bar{\nu}$, and $\Phi_{bc} \to \Phi_{cc} + \tau + \bar{\nu}_{\tau}$—are computed, yielding a total inclusive decay width of $\Gamma(\Xi_{bcq} \to \Xi_{ccq} + X) \approx 4.1 \times 10^{-13}$ GeV. We also evaluate the potential background from $B_c^- \to \Xi_{cc} + X$ decays, finding a rate of approximately $ 3.0 \times 10^{-14}$ GeV, which is roughly one order of magnitude smaller than the signal process. Our results suggest that the inclusive decay $\Xi_{bc} \to \Xi_{cc} + X$ , followed by the well-established $\Xi_{cc}$ decay chains, provides a viable discovery channel for the doubly heavy baryon $\Xi_{bc}$ at the LHC.
\end{abstract}
\maketitle
\section{Introduction}

In the framework of the Standard Model, weak decays of heavy-flavor hadrons provide an important avenue for exploring non-perturbative effects of strong interactions and precisely testing the unitarity of the Cabibbo-Kobayashi-Maskawa (CKM) matrix. In recent years, with the rapid development of heavy-flavor physics experiments at the Large Hadron Collider (LHC), particularly following the LHCb collaboration's discovery of the doubly charmed baryon $\Xi_{cc}^{++}$ (with a mass of approximately 3621.40 MeV) in 2017 \cite{LHCb:2017iph} and subsequent observations of its various decay modes \cite{LHCb:2018pcs,LHCb:2019qed,LHCb:2019ybf} , doubly heavy baryons (hadrons containing two heavy quarks) have become a forefront research topic in particle physics. In this context, studying the weak decay properties of the doubly heavy baryon $\Xi_{bcq}$ (where $q=u,d,s$ represents a light quark), which contains both a bottom and a charm quark, not only helps to refine the theoretical framework of heavy-hadron spectroscopy but also provides crucial theoretical guidance for experimental searches for yet-unobserved members of the $\Xi_{bc}$ family.

At the quark level, the decay process $\Xi_{bcq} \to \Xi_{ccq} + X$ is essentially dominated by the $b \to c$ transition, with the low-energy effective Hamiltonian given by:
\[
\mathcal{H}_{\text{eff}} = \frac{G_F}{\sqrt{2}} V_{cb} \left[ \bar{c}\gamma^\mu(1-\gamma_5)b \right] \left[ \bar{q}'\gamma_\mu(1-\gamma_5)q'' \right] + \text{h.c.},
\]
where $G_F$ is the Fermi constant, $V_{cb}$ is the CKM matrix element, and $X$ denotes a lepton pair $\ell\bar{\nu}_\ell$ (semileptonic decay) or light hadrons (nonleptonic decay). This process involves the transition from the doubly heavy initial state ($bcq$) to the doubly charmed final state ($ccq$), offering a unique testing ground for the application of Heavy Quark Effective Theory (HQET) and perturbative QCD. Particularly in the limit where $m_b, m_c \gg \Lambda_{\text{QCD}}$, the independent heavy quark spin symmetry allows all $\Xi_{bc}^{('*)} \to \Xi_{cc}^{(*)}$ transitions to be described by a single form factor, greatly simplifying theoretical calculations and enhancing the reliability of predictions\cite{Flynn:2007qt}.

Theoretically, studying the decay $\Xi_{bcq} \to \Xi_{ccq} + X$ is motivated by multiple physics considerations. First, through precise calculations of the form factors and decay widths for this process, one can extract or verify the CKM matrix element $|V_{cb}|$, which is a crucial component in testing the unitarity of the third row of the CKM matrix in the Standard Model. Second, the doubly heavy baryon system contains two heavy quarks and one light quark, and its weak decay involves complex non-perturbative QCD effects, such as the dynamics of the light-quark cloud, final-state interactions, and factorization-breaking effects. By comparing calculations from different theoretical frameworks—including QCD sum rules, the Light-Front Quark Model (LFQM), and Lattice QCD—one can deepen the understanding of baryon internal structure\cite{Patel:2024mfn}. Furthermore, since both the initial and final states contain two heavy quarks, this decay provides an ideal laboratory for studying SU(3) flavor symmetry and its breaking effects in double-heavy physics\cite{Huang:2021jxt}.

Experimentally, although the $\Xi_{cc}^{++}$ has been discovered and its lifetime and several decay modes have been measured, the $\Xi_{bc}$ (containing $bcu$, $bcd$, and $bcs$ quark content) has not yet been experimentally confirmed. Theoretical studies indicate that $\Xi_{bc}$ baryons primarily decay via the $b$ quark (with a lifetime on the order of $10^{-13}$ seconds), and their semileptonic branching fractions can reach the percent to per mille level, which falls within the detection sensitivity of experiments such as LHCb\cite{Li:2017ndo}. Therefore, precise calculations of the decay rates and angular distributions for $\Xi_{bcq} \to \Xi_{ccq} + \ell\bar{\nu}_\ell$ provide direct guidance for setting optimized search conditions, identifying signal events, and analyzing backgrounds in experimental searches.

However, the theoretical description of this decay process faces numerous challenges. In semileptonic decays, the hadronic matrix element $\langle \Xi_{ccq} | \bar{c}\gamma^\mu(1-\gamma_5)b | \Xi_{bcq} \rangle$ is parameterized by six independent form factors, which are functions of the momentum transfer $q^2$ and encode non-perturbative information about the baryon wave functions. Currently, different theoretical approaches (such as the light-front quark model, covariant confined quark model, and QCD sum rules) yield somewhat different results for these form factors, primarily due to different treatments of the light-quark degrees of freedom and model dependencies\cite{Lu:2023rmq}. In the case of nonleptonic decays, the applicability of the factorization assumption, contributions from $W$-exchange diagrams, and rescattering effects between final-state hadrons (such as through intermediate states like $\Lambda_c^+ K^- \pi^+ \pi^+$) further increase the uncertainties in theoretical predictions\cite{Liu:2025hbf}.

In summary, systematic studies of the inclusive decay $\Xi_{bcq} \to \Xi_{ccq} + X$ are not only fundamental for refining the theory of weak decays of doubly heavy baryons but also serve as a critical link connecting precision tests of the Standard Model with experimental observations. This work aims to employ Non-relativistic Potential Quark Model to calculate the transition form factors for this process, thereby predicting observable quantities such as decay widths,  and branching fractions. These results will provide theoretical support for experiments such as LHCb in their search for $\Xi_{bc}$ baryons and explore the physical implications ranging from heavy quark symmetry to flavor SU(3) symmetry breaking.

The rest of the paper is organized as follows. In section 2, we derive the expression for the differential decay width through theoretical derivation. In section 3, We obtain the wave function by solving the non-relativistic Schrödinger equation, and then substitute it into the differential decay width expression. After numerical integration, we obtain the total decay width.  We discuss background contributions in Section 4 and present our conclusions in Section  5.

\section{ THEORETICAL FRAMEWORK AND DECAY RATES}

 We work in the heavy quark limit where both the bottom and charm masses significantly exceed the non-perturbative QCD scale, $m_{b,c} \gg \Lambda_{\rm QCD}\approx 200$ MeV. In this kinematical regime, the two heavy quarks inside $\Xi_{bcq}$ bind into a compact color-anti-triplet diquark—denoted $\Phi_{bc}$—whose spatial extent is parametrically smaller than the confinement radius $1/\Lambda_{\rm QCD}$. The resulting $\Xi_{bcq}$ (or $\Xi_{ccq}$) clustering constitutes a non-relativistic bound system, allowing us to treat the relative motion between the diquark and the anti-meson in the $v\ll c$ approximation. Consequently, the inclusive decay $\Gamma( \Xi_{bcq} \to  \Xi_{ccq} +X)$ is effectively described by the underlying quark transition $ \Phi_{bc} \to  \Phi_{cc} +X$, while the light quark $q$ remains a passive spectator throughout the decay process. The calculation of $ \Xi_{bcq} \to  \Xi_{ccq} +X$ can be decomposed into four channels, $ \Phi_{bc} \to  \Phi_{cc} +\bar{c} +s(d)$, $ \Phi_{bc} \to  \Phi_{cc} +\bar{u} +s(d)$, $ \Phi_{bc} \to  \Phi_{cc} +e(\mu) +\bar{v_e}(\bar{v_{\mu}})$, $ \Phi_{bc} \to  \Phi_{cc} +\tau +\bar{v_{\tau}}$. The calculation is carried out in the rest frame of the decaying beauty-charmed diquark state $\Phi_{b c}$. Due to the Pauli exclusion principle, the spin wave function of diquark $\Phi_{c c}$ can only be symmetrical, so the spin of diquark $\Phi_{c c}$ is 1. We further assume that the relative momentum within the bound state is non-relativistic. The diquark state vectors are defined as:
\begin{eqnarray}
\left|\Phi_{b c}(\mathbf{p}, \gamma,s,m_s)\right\rangle&=&\sqrt{ E_{bc}(p)}\int \frac{d^{3} p_1}{(2 \pi)^{3}} \tilde{\psi}_{\Phi_{b c}}(\mathbf{p_1}) C_{s_{1} s_{2}}^{s, m_{s}}\epsilon^{\gamma \alpha \beta}\left|b(\frac{m_{b} \mathbf{p}}{m_{bc}}+\mathbf{p_1}, \alpha, s_{1}) c(\frac{m_{c} \mathbf{p}}{m_{bc}}-\mathbf{p_1}, \beta, s_{2})\right\rangle,\nonumber\\   
\left|\Phi_{cc}(\mathbf{k}, \gamma,1,m)\right\rangle&=&\frac{\sqrt{2 E_{c c}(k)}}{2}  \int \frac{d^{3} p_2}{(2 \pi)^{3}} \tilde{\psi}_{\Phi_{c c}}(\mathbf{p_2})  C_{s_{1} s_{2}}^{1, m}\epsilon^{\gamma \alpha \beta}\left|c\left(\frac{\mathbf{k}}{2}+\mathbf{p_2}, \alpha, s_{1}\right) c\left(\frac{\mathbf{k}}{2}-\mathbf{p_2}, \beta, s_{2}\right)\right\rangle,
\end{eqnarray}
where   the four indices represent respectively the momentum with respect to laboratory frame of reference , colour indice , spin , the projection of spin in the z direction and repeated indices are summed over. The $\tilde{\psi}_{\Phi_{b c}}(\mathbf{p_1})$ and $\tilde{\psi}_{\Phi_{c c}}(\mathbf{p_2})$ represent the wave functions describing the relative momentum of quarks within the bound state. In the non-relativistic approximation, they are significant only when the momentum $\mathbf{p_1},\mathbf{p_2}$ is much smaller than the masses of the bound quarks $m_{bc},m_{cc}$. The bound states have been normalized such that
 \begin{eqnarray}
\left\langle \Phi_{b c}\left(\mathbf{p'}, \gamma',s',m'_s\right)|\Phi_{b c}\left(\mathbf{p}, \gamma,s,m_s\right)\right\rangle&=&2E_{bc}(p)(2\pi)^3\delta^3(\mathbf{p'-p})\delta_{\gamma'\gamma}\delta_{s's}\delta_{m'_{s}m_{s}},\nonumber\\
\left\langle \Phi_{c c}(\mathbf{k}', \gamma', m')|\Phi_{c c}(\mathbf{k}, \gamma, m)\right\rangle  &=&2E_{c c}(k)(2\pi)^3\delta^3(\mathbf{k'-k})\delta_{\gamma'\gamma}\delta_{m'm} .
\end{eqnarray}

The effective Hamiltonian involved in the $b\to c+\bar{c}+s$ processes is
\begin{eqnarray}\label{eq:effH}	
	&\mathcal{H}_{\text{eff}}=\frac{4 G_{F}}{\sqrt{2}} V_{c s}^{*} V_{c b}\left[C_{1} O_{1}+C_{2} O_{2}\right], \nonumber\\
&O_{1}=\left[\bar{c}_{\rho} \gamma^{\mu} P_{L} b_{\rho}\right]\left[\bar{s}_{ \lambda  } \gamma_{\mu} P_{L} c_{ \lambda}\right]\; , \ O_{2}=\left[\bar{c}_{\lambda} \gamma^{\mu} P_{L} b_{\rho}\right]\left[\bar{s}_{\rho} \gamma_{\mu} P_{L} c_{\lambda}\right] \; , 
\end{eqnarray}
where $P_{L}$ are the left-handed projectors. The operators $O_{1},O_{2} $ and coefficients $C_{1},C_{2}$ are evaluated at a subtraction point corresponding to the b-quark mass. The invariant matrix element for the decay is
\begin{eqnarray}\label{iM}	
	i\mathcal{M}
	=\frac{4 G_{F}}{\sqrt{2}} V_{c s}^{*} V_{c b}\left[C_{1} \left\langle \Phi_{c c}\bar{c}s|O_{1}| \Phi_{b c}\right\rangle+C_{2} \left \langle \Phi_{c c}\bar{c}s|O_{2}| \Phi_{b c}\right\rangle \right] ,
\end{eqnarray}
where
\begin{eqnarray}
\left\langle \Phi_{c c}\bar{c}s|O_{1}| \Phi_{b c}\right\rangle   
=&\frac{\sqrt{2E_{bc}(p) E_{c c}(k)}}{2} \int \frac{d^{3} p_2}{(2 \pi)^{3}} \tilde{\psi^{*}}_{\Phi_{c c}}(\mathbf{p_2}) C_{s'_{1} s'_{2}}^{s, m*}\epsilon^{\gamma' \alpha' \beta'}  \int \frac{d^{3} p_1}{(2 \pi)^{3}}  \tilde{\psi}_{\Phi_{b c}}(\mathbf{p_1}) C_{s_{1} s_{2}}^{1, m_{s}}\epsilon^{\gamma \alpha \beta} \nonumber\\
&\left \langle  c\left(\frac{\mathbf{k}}{2}+\mathbf{p_2}, \alpha', s'_{1}\right) c(\frac{\mathbf{k}}{2}-\mathbf{p_2}, \beta', s_{2}')     \bar{c}(p_{ \bar{c}},\theta,s_{ \bar{c}}) s(p_{s},\eta,s_{s}) \right|\nonumber\\
& \left[\bar{c}_{\rho} \gamma^{\mu} P_{L} b_{\rho}\right]\left[\bar{s}_{ \lambda  } \gamma_{\mu} P_{L} c_{ \lambda}\right]
\left |b(\frac{m_{b} \mathbf{p}}{m_{bc}}+\mathbf{p_1}, \alpha, s_{1}) c(\frac{m_{c} \mathbf{p}}{m_{bc}}-\mathbf{p_1}, \beta, s_{2}) \right\rangle ,	\label{cc1bc}\\
\left\langle \Phi_{c c}\bar{c}s|O_{2}| \Phi_{b c}\right\rangle   
=&\frac{\sqrt{2E_{bc}(p) E_{c c}(k)}}{2} \int \frac{d^{3} p_2}{(2 \pi)^{3}} \tilde{\psi^{*}}_{\Phi_{c c}}(\mathbf{p_2}) C_{s'_{1} s'_{2}}^{s, m*}\epsilon^{\gamma'  \beta'\alpha'}  \int \frac{d^{3} p_1}{(2 \pi)^{3}}  \tilde{\psi}_{\Phi_{b c}}(\mathbf{p_1}) C_{s_{1} s_{2}}^{1, m_{s}}\epsilon^{\gamma \alpha \beta} \nonumber\\
&\left \langle  c\left(\frac{\mathbf{k}}{2}+\mathbf{p_2}, \alpha', s'_{2}\right) c(\frac{\mathbf{k}}{2}-\mathbf{p_2}, \beta', s_{1}')     \bar{c}(p_{ \bar{c}},\theta,s_{ \bar{c}}) s(p_{s},\eta,s_{s}) \right|\nonumber\\
& \left[\bar{c}_{\rho} \gamma^{\mu} P_{L} b_{\rho}\right]\left[\bar{s}_{ \lambda  } \gamma_{\mu} P_{L} c_{ \lambda}\right]
\left |b(\frac{m_{b} \mathbf{p}}{m_{bc}}+\mathbf{p_1}, \alpha, s_{1}) c(\frac{m_{c} \mathbf{p}}{m_{bc}}-\mathbf{p_1}, \beta, s_{2}) \right\rangle .\label{cc2bc}
\end{eqnarray}


After some calculation , eq. (\ref{cc1bc}) and (\ref{cc2bc}) becomes

\begin{eqnarray}\label{cc1bc1}	
\left\langle \Phi_{c c}\bar{c}s|O_{1}| \Phi_{b c}\right\rangle=&\frac{\sqrt{E_{bc}(0) E_{c c}(k)}}{\sqrt{2E_c(\mathbf{p_1}+\mathbf{k})E_b(\mathbf{p_1})}} \int \frac{d^{3} p_1}{(2 \pi)^{3}} \tilde{\psi^{*}}_{\Phi_{c c}}(|\mathbf{p_1}+\frac{ \mathbf{k}}{2}|) (\frac{C_{s'_{1} s_{2}}^{s, m*}+C_{ s_{2}s'_{1}}^{s, m*}}{2})  \tilde{\psi}_{\Phi_{b c}}(\mathbf{p_1})  C_{s_{1} s_{2}}^{1, m_s} 2\delta_{\gamma' \gamma}\delta_{\theta\eta }\nonumber\\
&\left[\bar{u}^{(s)}\left(\mathbf{p}_{\mathbf{s}},s_{s}\right) \gamma^{\mu} P_{L} v^{(\bar{c})}\left(\mathbf{p}_{\bar{c}}, s_{\bar{c}}\right)\right]\left[\bar{u}^{(c)}\left(\mathbf{p_1}+\mathbf{k}, s_{1}^{\prime}\right) \gamma_{\mu} P_{L} u^{(b)}\left(\frac{m_{b} \mathbf{p}}{m_{bc}}+\mathbf{p_1}, s_{1}\right)\right],
	\end{eqnarray}
\begin{eqnarray}\label{cc2bc1}	
\left \langle \Phi_{c c}\bar{c}s|O_{2}| \Phi_{b c}\right\rangle 
=&\frac{\sqrt{E_{bc}(0) E_{c c}(k)}}{\sqrt{2E_c(\mathbf{p_1}+\mathbf{k})E_b(\mathbf{p_1})}} \int \frac{d^{3} p_1}{(2 \pi)^{3}} \tilde{\psi^{*}}_{\Phi_{c c}}(|\mathbf{p_1}+ \frac{ \mathbf{k}}{2}|) (\frac{C_{s'_{1} s_{2}}^{s, m*}+C_{ s_{2}s'_{1}}^{s, m*}}{2})  \tilde{\psi}_{\Phi_{c c}}(\mathbf{p_1})  C_{s_{1} s_{2}}^{1,m_{s} } (\delta_{\gamma' \gamma}\delta_{\theta\eta }-\delta_{\gamma' \eta}\delta_{\gamma\theta }) \nonumber\\
&\left[\bar{u}^{(s)}\left(\mathbf{p}_{\mathbf{s}},s_{s}\right) \gamma^{\mu} P_{L} v^{(\bar{c})}\left(\mathbf{p}_{\bar{c}}, s_{\bar{c}}\right)\right]
\left[\bar{u}^{(c)}\left(\mathbf{p_1}+\mathbf{k}, s_{1}^{\prime}\right) \gamma_{\mu} P_{L} u^{(b)}\left(\frac{m_{b} \mathbf{p}}{m_{bc}}+\mathbf{p_1}, s_{1}\right)\right].
\end{eqnarray}

In addition, bottom quark momentum can be approximated as $\frac{m_{b} \mathbf{p}}{m_{bc}}$ , the wave function restricts $|\mathbf{p_1}+\mathbf{k}|$ to be much less than the charm quark mass, which means we ca make the replacement $|\mathbf{p_1}+\mathbf{k}|\to \frac{\mathbf{k}}{2}$ in $E_c$, so 
$ u^{(b)}(\frac{m_{b} \mathbf{p}}{m_{bc}}+\mathbf{p_1}, s_1 ) \to u^{(b)} (\frac{m_{b} \mathbf{p}}{m_{bc}}, s_1 ),\bar{u}^{(c)}\left(\mathbf{p_1}+\mathbf{k}, s_{1}^{\prime}\right)\to\bar{u}^{(c)}\left(\frac{\mathbf{k}}{2}, s_{1}^{\prime}\right)$ . In the non-relativistic limit, the masses of the bound states are approximately equal to the sum of their constituent quark masses, we can
set $\frac{E_{bc}}{E_{b}}=\frac{m_{bc}}{m_{b}},\frac{E_{cc}}{E_{c}}=\frac{m_{cc}}{m_{c}}=2$. After making these replacements, eq. (\ref{iM}) becomes
\begin{eqnarray}	
i\mathcal{M} \simeq& \frac{4 G_{F}}{\sqrt{2}} V_{c s}^{*} V_{c b}\left(2C_{1}\delta_{\gamma' \gamma}\delta_{\theta\eta } +C_{2}(\delta_{\gamma' \gamma}\delta_{\theta\eta }-\delta_{\gamma' \eta}\delta_{\gamma\theta })\right) \sqrt{\frac{\left(m_{b}+m_{c}\right)}{ m_{b}}}   C_{s_{1}, s_{2}}^{1, m_{s} } \frac{\mathcal{I}\left(\frac{k}{2} \right)}{2}\nonumber\\
&\times (C_{s'_{1} s_{2}}^{s, m*}\left[\bar{u}^{(s)}\left(\mathbf{p}_{\mathbf{s}}, s_{s}\right) \gamma^{\mu} P_{L} v^{(\bar{c})}\left(\mathbf{p}_{\bar{c}}, s_{\bar{c}}\right)\right]\left[\bar{u}^{(c)}\left(\frac{\mathbf{k}}{2} , s_{1}^{\prime}\right) \gamma_{\mu} P_{L} u^{(b)}\left(\frac{m_{b} \mathbf{p}}{m_{bc}}, s_{1}\right)\right]\nonumber\\
&+C_{ s_{2}s'_{2}}^{s, m*}\left[\bar{u}^{(s)}\left(\mathbf{p}_{\mathbf{s}}, s_{s}\right) \gamma^{\mu} P_{L} v^{(\bar{c})}\left(\mathbf{p}_{\bar{c}}, s_{\bar{c}}\right)\right]\left[\bar{u}^{(c)}\left(\frac{\mathbf{k}}{2} , s_{2}^{\prime}\right) \gamma_{\mu} P_{L} u^{(b)}\left(\frac{m_{b} \mathbf{p}}{m_{bc}}, s_{1}\right)\right]),
\end{eqnarray}
the overlap integral of the wave functions defines the form factor:
\begin{eqnarray}	\label{Ik}	
 \mathcal{I}(k)=&\int \frac{d^{3} p_1}{(2 \pi)^{3}} \tilde{\psi}_{\Phi_{c c}}^{*}(|\mathbf{p_1}+\mathbf{k}|) \tilde{\psi}_{\Phi_{b c}}(p_1)
=4 \pi \int d r r^{2} \psi_{\Phi_{c c}}^{*}(r) \psi_{\Phi_{b c}}(r) \frac{\sin (k r)}{k r}.
\end{eqnarray}

After the amplitude modulus square, we take the average of the initial state and sum the final state. The part related to color indicators is
\begin{eqnarray}
	\frac{1}{3}\sum_{\gamma'\gamma  \eta \theta}|2C_{1}\delta_{\gamma' \gamma}\delta_{\theta\eta } +C_{2}(\delta_{\gamma' \gamma}\delta_{\theta\eta }-\delta_{\gamma' \eta}\delta_{\gamma\theta })|^{2}
	=&4[3C_1^2+C_2 ^2+2C_1C_2 ].  
\end{eqnarray}
The remaining part is
\begin{eqnarray}
	&\frac{1}{3}\sum_{s,m,m_s,s_s,s_{\bar{c}}}|C_{s_{1}, s_{2}}^{1, m_s }C_{s'_{1}, s_{2}}^{s,m *}
	[\bar{u}^{(s)}\left(\mathbf{p}_{\mathbf{s}}, s_{s}\right) \gamma^{\mu} (1-\gamma_{5}) v^{(\bar{c})}\left(\mathbf{p}_{\bar{c}}, s_{\bar{c}}\right)[\bar{u}^{(c)}\left(\frac{\mathbf{k}}{2} , s_{1}^{\prime}\right) \gamma_{\mu} (1-\gamma_{5}) u^{(b)}\left(\frac{m_{b} \mathbf{p}}{m_{bc}}, s_{1}\right)]|^2\nonumber\\ 
	=&128(\frac{m_{b}p}{m_{bc}}\cdot p_{\bar{c}} )(p_s\cdot \frac{{k}}{2})\\
	&\frac{1}{3}\sum_{s,m,m_s,s_s,s_{\bar{c}}}|C_{s_{1}, s_{2}}^{1, m_s }C_{, s_{2}s'_{2}}^{s,m *}
	[\bar{u}^{(s)}\left(\mathbf{p}_{\mathbf{s}}, s_{s}\right) \gamma^{\mu} (1-\gamma_{5}) v^{(\bar{c})}\left(\mathbf{p}_{\bar{c}}, s_{\bar{c}}\right)[\bar{u}^{(c)}\left(\frac{\mathbf{k}}{2} , s_{2}^{\prime}\right) \gamma_{\mu} (1-\gamma_{5}) u^{(b)}\left(\frac{m_{b} \mathbf{p}}{m_{bc}}, s_{1}\right)]|^2\nonumber\\
	=&128(\frac{m_{b}p}{m_{bc}}\cdot p_{\bar{c}} )(p_s\cdot \frac{{k}}{2})
\end{eqnarray}
The cross term is 0. The result after  amplitude modulus square is
\begin{eqnarray}
	&\frac{1}{3}\sum_{m,s,m_s,s_s,s_c}\frac{1}{3}\sum_{\gamma'  \gamma\eta \theta}|\mathcal{M}|^{2}		
	=\frac{256G_{F}^2m_{bc}| V_{c s}^{*} V_{c b}|^2}{m_b}(3C_1^2+C_2 ^2+2C_1C_2)| \mathcal{I}\left(\frac{k}{2} \right)|^2  (\frac{m_{b} p}{m_{bc}}\cdot p_{\bar{c}} )(p_s\cdot \frac{{k}}{2}).
\end{eqnarray}

Substituting it into the three-body decay width formula yields
\begin{eqnarray}
	\Gamma
	=&\frac{1}{16(2 \pi)^{5}m_{b c}}\int  \frac{d^{3} \vec{p}_{k} d^{3} \vec{p}_{s} d^{3} \vec{p}_{{\bar{c}}}}{ E_{k} E_{s} E_{{\bar{c}}}}\delta^{4}\left(k+p_s+p_{\bar{c}}-p\right)\frac{256G_{F}^2m_{bc}| V_{c s}^{*} V_{c b}|^2}{m_b}\nonumber\\ 
	&(3C_1^2+C_2 ^2+2C_1C_2)|\mathcal{I}\left(\frac{k}{2} \right)|^2  (\frac{m_{b}p}{m_{bc}}\cdot p_{\bar{c}} )(p_s\cdot \frac{{k}}{2}).
\end{eqnarray}

The integral is obtained by dividing all phase spaces except energy	$E_k$, the differential decay width of energy $E_k$ is
\begin{eqnarray}\label{dGdE}	
\frac{d \Gamma}{d E_{k}}=&\left(\frac{G_{F}^{2}}{2\pi^{3}m_{bc}}\right)\left(3C_1^2+C_2 ^2+2C_1C_2\right)\left| V_{c s}^*V_{c b}\right|^{2}\left|\mathcal{I}(\frac{k}{2} )\right|^{2}\nonumber\\ &[\frac{-s_{s,cc}^3}{6m_{b c}}+\frac{s_{s,cc}^2(m_{bc}^{2}+m_{\bar{c}}^2+m_{cc}^{2})}{4m_{b c}}-\frac{s_{s,cc}(m_{bc}^{2}+m_{\bar{c}}^2)m_{cc}^{2}}{2m_{b c}}]|_{\min }^{\max }
\end{eqnarray}
where $s_{s,cc}$ is finial state $\Phi_{c c}$ and $s$ invariant mass. The maximum and minimum values are
\begin{eqnarray}
\left(s_{s,cc}\right)_{\max }=\left(E_{s}^{*}+E_{cc}^{*}\right)^{2}-\left(\sqrt{E_{s}^{* 2}-m_{s}^{2}}-\sqrt{E_{cc}^{* 2}-m_{cc}^{2}}\right)^{2}, \nonumber\\
\left(s_{s,cc}\right)_{\min }= \left(E_{s}^{*}+E_{cc}^{*}\right)^{2}-\left(\sqrt{E_{s}^{* 2}-m_{s}^{2}}+\sqrt{E_{cc}^{* 2}-m_{cc}^{2}}\right)^{2},
\end{eqnarray}
where $ E_{s}^{*},  E_{cc}^{*}$ is the energy of $\Phi_{c c}$ and $s$ in the stationary system of $\Phi_{c c},s$ particles,
\begin{eqnarray}
E_{s}^{*}=\left(m_{cs}^{2}-m_{c}^{2}+m_{s}^{2}\right) / 2 m_{cs} , E_{cc}^{*}=\left(m_{bc}^{2}-m_{cs}^{2}-m_{cc}^{2}\right) / 2 m_{cs}.
\end{eqnarray}

\section{Numerical analysis}
The evaluation of the form factors requires knowledge of the bound-state wave functions $\psi_{\Phi_{b c}}$ $\psi_{\Phi_{c c}}$ in position space. Within the non-relativistic framework, the momentum-space wave function is obtained by first solving the Schrödinger equation with the appropriate inter-quark potential and then Fourier-transforming the resulting coordinate-space wave function. We use Cornell potentials as inter-quark potential:
\begin{eqnarray}
V_{\Phi_{cc}}(r)=-\frac{2}{3}\left(\frac{0.5}{r}\right)+\frac{1}{2}*0.2  r, \quad V_{\Phi_{bc}}(r)=-\frac{2}{3}\left(\frac{0.4}{r}\right)+\frac{1}{2}*0.2  r . 
\end{eqnarray}

We employ the Gaussian expansion method to approximate the true physical state within a finite basis space, as described in \cite{caoluchenhong}. This approach involves three parameters: \( r_1 \), \( r_{\text{max}} \), and \( N_{\text{max}} \). When \( r_1 = 0.1 \), increasing \( r_{\text{max}} \) and \( N_{\text{max}} \) leads to convergence of the mean-square radius, yielding \( \langle r^2 \rangle_{\Phi_{cc}} = 7.4 \) and \( \langle r^2 \rangle_{\Phi_{bc}} = 5.5 \).

The exact wave function typically involves a large number of basis terms, and the corresponding expression for \( \mathcal{I}(k) \) becomes even more complex. Using such accurate representations would significantly slow down the subsequent numerical integration. Since the current work aims at an order-of-magnitude estimate rather than high precision, we adopt the wave functions obtained from a simple Coulomb potential as a reasonable approximation, which greatly simplifies the evaluation of the integral.


It turns out that the Coulomb-like wave functions
 \begin{eqnarray}\label{bccc}
	\psi_{\Phi_{b c}}(r) &=&\frac{1}{\sqrt{\pi}}\left(\frac{3}{\left\langle r^{2}\right\rangle_{\Phi_{bc}}}\right)^{3 / 4} Exp\left(-\frac{\sqrt{3} r}{\sqrt{\left\langle r^{2}\right\rangle_{\Phi_{bc}}}}\right) ,\nonumber\\
	\psi_{\Phi_{c c}}(r) &=&\frac{1}{\sqrt{\pi}}\left(\frac{3}{\left\langle r^{2}\right\rangle_{\Phi_{c c}}}\right)^{3 / 4} Exp\left(-\frac{\sqrt{3} r}{\sqrt{\left\langle r^{2}\right\rangle_{\Phi_{c c}}}}\right)	.	
\end{eqnarray} 
 are good approximations to thenumerical ones. Plugging eq.(\ref{bccc}) into eq.(\ref{Ik}) yields a compact analytic approximation for the form factor,
\begin{eqnarray}
\mathcal{I}(k) &=\frac{72(\sqrt{\left\langle r^{2}\right\rangle_{\Phi_{bc}}}+\sqrt{\left\langle r^{2}\right\rangle_{\Phi_{cc}}})(\left\langle r^{2}\right\rangle_{\Phi_{bc}} {\left\langle r^{2}\right\rangle_{\Phi_{cc}}})^{3 / 4}}{\left(3 \left\langle r^{2}\right\rangle_{\Phi_{bc}}+3{\left\langle r^{2}\right\rangle_{\Phi_{cc}}}+6 \sqrt{{\left\langle r^{2}\right\rangle_{\Phi_{bc}} }{\left\langle r^{2}\right\rangle_{\Phi_{cc}}}}+{\left\langle r^{2}\right\rangle_{\Phi_{bc}}} {\left\langle r^{2}\right\rangle_{\Phi_{cc}}} k^{2}\right)^{2}}	
\end{eqnarray} 	
Based on the relationship between momentum and energy, we can make the following substitution
\begin{eqnarray}
\mathcal{I}(k/2)\to\mathcal{I}(E_k) &=\frac{72(\sqrt{\left\langle r^{2}\right\rangle_{\Phi_{bc}}}+\sqrt{\left\langle r^{2}\right\rangle_{\Phi_{cc}}})(\left\langle r^{2}\right\rangle_{\Phi_{bc}} {\left\langle r^{2}\right\rangle_{\Phi_{cc}}})^{3 / 4}}{\left(3 \left\langle r^{2}\right\rangle_{\Phi_{bc}}+3{\left\langle r^{2}\right\rangle_{\Phi_{cc}}}+6 \sqrt{{\left\langle r^{2}\right\rangle_{\Phi_{bc}} }{\left\langle r^{2}\right\rangle_{\Phi_{cc}}}}+{\left\langle r^{2}\right\rangle_{\Phi_{bc}}} {\left\langle r^{2}\right\rangle_{\Phi_{cc}}} \frac{E_k^2-m_{cc}^2}{4}\right)^{2}}	
\end{eqnarray} 	
\begin{figure}
	\centering
	\includegraphics[width=0.5\linewidth]{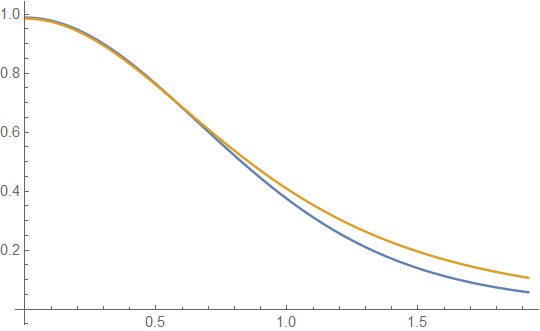}
	\caption{The form factor $\mathcal{I}(k)$ , defined in Eq.(\ref{Ik}), is evaluated using both the numerical ground-state wave function (blue) and the approximate form given in Eq.(\ref{bccc}) (yellow).}
	\label{fig:1}
\end{figure}

Based on the relation between momentum and energy, we substitute $k\to \sqrt{E_k^2-m_{cc}^2}$. Taking \(m_c = 1.5\,\text{GeV}\) and \(m_b = 4.5\,\text{GeV}\), the form factor \(I(k)\) obtained from the numerical ground-state wave function is displayed in Fig. 1. The energy of a diquark $\Phi_{cc}$ is
\begin{eqnarray}	
	&E_{k}=\frac{m_{b c}^2+m_{cc}^2-m_{{\bar{c}}s}^2}{2m_{b c}},
\end{eqnarray}
where $m_{{\bar{c}}s}$ is finial state $\bar{c}$ and $s$ invariant mass, $m_{{\bar{c}}s}\ge m_{\bar{c}}+m_s,m_{\bar{c}}\gg m_s$. When the mass of $s$ quark is ignored, the mass of $bc$ are approximately equal to the sum of their constituent quark masses, the minimum value of diquark energy is 0, the maximum value of diquark energy is
\begin{eqnarray}	 
	&E_{k}|_{max}=\frac{m_{b c}^2+m_{cc}^2-m_{\bar{c}}^2}{2m_{b c}} \approx 3.56\text{GeV}.
\end{eqnarray}
Integrating over energy from 0 to the maximum, we find that the decay rate is
\begin{eqnarray}	
	&\Gamma(\Phi_{bc} \rightarrow \Phi_{cc}+\bar{c}+s(d))\approx4.7\times 10^{-14}\text{GeV}.
\end{eqnarray}

Similarly, for the decay channel $\Phi_{bc} \rightarrow \Phi_{cc}+\bar{u}+s(d) $, the differential width  is
\begin{eqnarray}
	\frac{d \Gamma}{d E_k}
	=&\frac{G_{F}^2| V_{u s}^{*} V_{c b}|^2\left(3C_1^2+C_2^2+2C_1C_2\right)}{\pi^{3}}| \mathcal{I}\left(E_k\right)|^2  \sqrt{E_k^2-m_{cc}^2} \frac{[3E_k(m_{bc}^2+m_{cc}^2)-2m_{bc}(2E_k^2+m_{cc}^2)]}{3},
\end{eqnarray}
the maximum value of diquark energy is
\begin{eqnarray}	
	&E_{k}|_{max}=\frac{m_{c c}^2+m_{b c}^2}{2m_{b c}}\approx 3.75\text{GeV}.
\end{eqnarray}
yielding the decay rate is
\begin{eqnarray}		
	& \Gamma(\Phi_{bc} \rightarrow \Phi_{cc}+\bar{u}+s(d))\approx2.1\times 10^{-13}\text{GeV}.
\end{eqnarray}

For the semileptonic channel , the differential width of the decay channel $\Phi_{bc} \rightarrow \Phi_{cc}+e(\mu)+\bar{v_e}(\bar{v_{\mu}})$ is
\begin{eqnarray}
	&\frac{d \Gamma}{d E_k}
	=\left(\frac{G_{F}^{2}}{ \pi^{3}}\right)2|V_{c b}|^{2}|\mathcal{I}\left(E_k\right)|^2  \sqrt{E_k^2-m_{cc}^2} \frac{[3E_k(m_{bc}^2+m_{cc}^2)-2m_{bc}(2E_k^2+m_{cc}^2)]}{3},
\end{eqnarray}
the maximum value of diquark energy is
\begin{eqnarray}	
	&E_{k}|_{max}=\frac{m_{c c}^2+m_{b c}^2}{2m_{b c}}\approx 3.75\text{GeV},
\end{eqnarray}
the decay rate is
\begin{eqnarray}		
	&\Gamma(\Phi_{bc} \rightarrow \Phi_{cc}+e(\mu)+\bar{v_e}(\bar{v_{\mu}}))\approx1.3\times 10^{-13}\text{GeV}.
\end{eqnarray}

For the differential width of the decay channel $\Phi_{bc} \rightarrow \Phi_{cc}+\tau+\bar{v_{\tau}}$ is
\begin{eqnarray}
	\frac{d \Gamma}{d E_k}
	=&\left(\frac{G_{F}^{2}}{ \pi^{3}}\right)|V_{c b}|^{2}|\mathcal{I}\left(E_k\right)|^2 [\frac{-s_{s,cc}^3}{6m_{bc}}+\frac{s_{s,cc}^2(m_{bc}^{2}+m_{\bar{c}}^2+m_{cc}^{2})}{4m_{b c}}-\frac{s_{s,cc}(m_{bc}^{2}+m_{\bar{c}}^2)m_{cc}^{2}}{2m_{b c}}]|_{\min }^{\max }	
\end{eqnarray}
The energy of a diquark $\Phi_{cc}$ is
\begin{eqnarray}
	E_{\Phi_{c c}}=\frac{m_{b c}^2+m_{c c}^2-m_{\tau \bar{v_{\tau}}}^2}{2m_{b c}},
\end{eqnarray}
where $m_{\tau \bar{v_{\tau}}}$ is finial state $\tau $ and $\bar{v_{\tau}}$ invariant mass,
	 $m_{\tau \bar{v_{\tau}}}\ge m_{\tau}+m_{\bar{v_{\tau}}},m_{\tau}\gg m_{\bar{v_{\tau}}} $.
the maximum value of diquark energy is
\begin{eqnarray}  
	&E_{k}|_{max}=\frac{m_{\Phi_{cc}}^2+m_{b c}^2-m_{\tau}^2}{2m_{b c}}\approx 3.48\text{GeV}.
\end{eqnarray}
yielding the decay rate is
\begin{eqnarray} 	
	& \Gamma(\Phi_{bc} \rightarrow \Phi_{cc}+\tau+\bar{v_{\tau}})\approx1.7\times 10^{-14}\text{GeV}.
\end{eqnarray}

Summing all channels, the total inclusive decay rate is
\begin{eqnarray}
	\Gamma(\Xi_{bcq} \rightarrow \Xi_{ccq}+X) 
	&=& \Gamma(\Phi_{bc} \rightarrow \Phi_{cc}+\bar{c}+s(d))+ \Gamma(\Phi_{bc} \rightarrow \Phi_{cc}+\bar{u}+s(d))+ \nonumber\\
	&&\Gamma(\Phi_{bc} \rightarrow \Phi_{cc}+e/\mu+\bar{v_e}(\bar{v_{\mu}}))+\Gamma(\Phi_{bc} \rightarrow \Phi_{cc}+\tau+\bar{v_{\tau}})\nonumber\\
	&\approx&4.1\times 10^{-13}\text{GeV}
\end{eqnarray}

Our calculation relies on the non-relativistic approximation for the heavy diquark system, where the relative momentum of constituent quarks is assumed to be small compared to their masses. The use of Coulomb-like wave functions as approximations to the exact numerical solutions introduces uncertainties at the factor-of-two level. Additionally, we have neglected contributions from radially or orbitally excited $\Phi_{cc}$ states, final-state interactions, and relativistic corrections, which may modify the results at the $20\%$–$30\%$ level.

 \section{BACKGROUND CONSIDERATIONS}
    Another process that generates the same final-state topology is $B_c\to \Xi_{cc}+X$, with the dominant quark-level transition $\bar{b}\to \bar{c}+c+\bar{s}$. Following a similar procedure, we define the initial and final states of decay
\begin{eqnarray}
	\left|\bar{B}_{c}\left(\mathbf{p}, s, m_{s}\right)\right\rangle&=&\frac{\sqrt{2 E_{\bar{B}_{c}}(p)}}{\sqrt{3}} \int \frac{d^{3} p_1}{(2 \pi)^{3}} \tilde{\psi}_{\bar{B}_{c}}(\mathbf{p_1}) C_{s_{1} s_{2}}^{s, m_{s}}\left|b(\frac{m_{b} \mathbf{p}}{m_{b}+m_{c}}+\mathbf{p_1}, \delta, s_{1}) \bar{c}(\frac{m_{c} \mathbf{p}}{m_{b}+m_{c}}-\mathbf{p_1}, \delta, s_{2})\right\rangle\nonumber\\
	\left|\Phi_{ \bar{c} \bar{c}}(\mathbf{k}, \gamma, m)\right\rangle&=&\frac{\sqrt{2 E_{\Phi_{ \bar{c} \bar{c} }}(k)}}{2}  \int \frac{d^{3} p_2}{(2 \pi)^{3}} \tilde{\psi}_{\Phi_{ \bar{c} \bar{c}}}(\mathbf{p_2}) \epsilon^{\gamma \alpha \beta} C_{s_{1} s_{2}}^{1, m}\left|\bar{c}\left(\frac{\mathbf{k}}{2}+\mathbf{p_2}, \alpha, s_{1}\right) \bar{c}\left(\frac{\mathbf{k}}{2}-\mathbf{p_2}, \beta, s_{2}\right)\right\rangle
\end{eqnarray}

The invariant matrix element for the decay is	
\begin{eqnarray}\label{M1}
	&\mathcal{M} =\frac{4 G_{F}}{\sqrt{6}} V_{c s}^{*} V_{c b}(C_{1}-C_{2}) \int\frac{d^{3} p_1}{(2 \pi)^{3}}  \frac{\sqrt{m_{\bar{B}_{c}} E_{\Phi_{\bar{c} \bar{c}}}(k)}}{\sqrt{E_{\bar{c}}(\mathbf{p_1}+\mathbf{k})E_b(\mathbf{p_1})}} \tilde{\psi}^{*}_{\Phi_{\bar{c} \bar{c}}}(\mathbf{p_1}+\frac{ \mathbf{k}}{2})\tilde{\psi}_{\bar{B}_{c}}(\mathbf{p_1}) \epsilon^{\gamma \theta\eta } \nonumber\\  
	& C_{s'_{2}, s_{2}}^{1, m *} C_{s_{1}, s_{2}}^{s, m_{s}} \left[\bar{u}^{(s)}\left(\mathbf{p}_{\mathbf{s}}, s_{s}\right) \gamma^{\mu} P_{L} v^{(\bar{c})}\left(\mathbf{p_1}+\mathbf{k}, s_{2}^{\prime}\right)\right]\left[\bar{u}^{(c)}\left(\mathbf{p}_{\mathbf{c}}, s_{c}\right) \gamma_{\mu} P_{L} u^{(b)}\left(\mathbf{p_1}, s_{1}\right)\right]
\end{eqnarray}      

After making same replacements, eq. (\ref{M1}) becomes
\begin{eqnarray}
\mathcal{M} \simeq& \frac{4 G_{F}}{\sqrt{6}} V_{c s}^{*} V_{c b}(C_{1}-C_{2}) \sqrt{\frac{2\left(m_{b}+m_{c}\right)}{ m_{b}}}  \epsilon^{\gamma \eta \theta } C_{s'_{2}, s_{2}}^{1, m *} C_{s_{1}, s_{2}}^{s, m_{s}} \mathcal{I}(\frac{\mathbf{k}}{2})\nonumber\\
&\left[\bar{u}^{(s)}\left(\mathbf{p}_{\mathbf{s}}, s_{s}\right) \gamma^{\mu} P_{L} v^{(\bar{c})}\left( \frac{\mathbf{k}}{2}, s_{2}^{\prime}\right)\right]\left[\bar{u}^{(c)}\left(\mathbf{p}_{\mathbf{c}}, s_{c}\right) \gamma_{\mu} P_{L} u^{(b)}\left(\frac{m_{b} \mathbf{p}}{m_{bc}}, s_{1}\right)\right]
\end{eqnarray}
 The result after  amplitude modulus square is
\begin{eqnarray}
	&\frac{1}{4}\sum_{m,s,m_s,s_s,s_c}\sum_{\gamma \theta \eta }|\mathcal{M}|^{2}		
	=\frac{256m_{bc}}{m_b}G_{F}^2| V_{c s}^{*} V_{c b}|^2(C_{1}-C_{2})^2| \mathcal{I}\left( \frac{\mathbf{k}}{2}\right)|^2 (\frac{m_{b}p}{m_{bc}}\cdot \frac{k}{2})(p_s\cdot p_c)
\end{eqnarray}
Substituting it into the three-body decay width formula yields
\begin{eqnarray}
	\Gamma
	=&\frac{1}{16(2 \pi)^{5}m_{b c}}\int  \frac{d^{3} \vec{p}_{k} d^{3} \vec{p}_{s} d^{3} \vec{p}_{c}}{ E_{k} E_{s} E_{c}}\delta^{4}\left(k+p_s+p_c-p\right)\frac{256G_{F}^2m_{bc}| V_{c s}^{*} V_{c b}|^2}{m_b}\nonumber\\ 
	&(C_{1}-C_{2})^2|\mathcal{I}\left(\frac{k}{2} \right)|^2 (\frac{m_{b}p}{m_{bc}}\cdot\frac{k}{2})(p_s\cdot p_c)
\end{eqnarray}

The integral is obtained by dividing all phase spaces except energy	$k$
\begin{eqnarray}
\frac{d \Gamma}{d k}	
	=& \frac{G_{F}^{2}}{ \pi^{3}}(C_{1}-C_{2})^2\left| V_{c s}^*V_{c b}\right|^{2}\left|\mathcal{I}(\frac{k}{2} )\right|^{2} k^{2} \frac{\left(m_{b c}^{2}+m_{ \Phi_{\bar{c} \bar{c}}}^{2}-m_{c}^{2}-2 E_{ \Phi_{\bar{c} \bar{c}}}(k) m_{b c}\right)^{2}}{\left(m_{bc}^{2}+m_{ \Phi_{\bar{c} \bar{c}}}^{2}-2 E_{ \Phi_{\bar{c} \bar{c}}}(k) m_{b c}\right)}
\end{eqnarray}
the decay rate is	
\begin{eqnarray}
	 \Gamma(\bar{B}_{c}\rightarrow \Phi_{\bar{c} \bar{c}}+c+s) 
	\approx&3.0\times 10^{-14}\text{GeV},
\end{eqnarray}
which is roughly one order of magnitude smaller than the signal process. This suggests that the inclusive $\Xi_{bc} \to \Xi_{cc} + X$ decay provides a viable discovery channel for the doubly heavy baryon $\Xi_{bc}$.
\section{Conclusion and Outlook}

In this work, we have investigated the inclusive weak decay $\Xi_{bcq} \to \Xi_{ccq} + X$ within the framework of the non-relativistic potential quark model. By treating the heavy diquarks $\Phi_{bc}$ and $\Phi_{cc}$ as compact color anti-triplet bound states, we calculated the transition form factors through the overlap integral of the Schrödinger wave functions obtained from the Cornell potential.The total inclusive decay rate is estimated to be:
$\Gamma(\Xi_{bcq} \to \Xi_{ccq} + X) \approx 4.1 \times 10^{-13} \text{ GeV}.$ We also evaluated the potential background from $B_c^- \to \Xi_{cc} + X$ decays, finding a rate of approximately $ 3.0 \times 10^{-14}$ GeV.

The observation of $\Xi_{bc}$ baryons remains one of the important goals in heavy-flavor physics. The significant branching ratio obtained in this work supports the proposal that inclusive decays to $\Xi_{cc}$ final states, followed by the well-established $\Xi_{cc}$ decay chains (e.g., $\Xi_{cc}^{++} \to \Lambda_c^+ K^- \pi^+ \pi^+$), provide a promising strategy for experimental detection. Future improvements could incorporate the full relativistic treatment of the light spectator quark, lattice QCD determinations of the form factors, and more sophisticated handling of the nonleptonic final-state interactions.

In summary, this study provides a quantitative theoretical foundation for the experimental search for $\Xi_{bc}$ baryons and highlights the importance of inclusive decay modes in exploring the spectroscopy of doubly heavy hadrons.

{\it Acknowledgement.} --- This work is supported by the School Research Fund under  Grant No. 1411005301. 


\end{document}